\documentclass[twoside]{ilcws08}
\usepackage[latin1]{inputenc}
\usepackage[dvips]{graphicx,epsfig,color}
\usepackage{wrapfig,rotating,units}
\usepackage{amssymb,amsmath,array}

\begin{document}
\title{First Results from New High Magnetic Field Measurements with the MediTPC Prototype} 

\author{Ralf Diener$^{1,3}$ and Diana Linzmaier$^{2,3}$
\vspace{.3cm}\\
1- University of Hamburg, Physics Department, Institut f{\"u}r Experimentalphysik\\
Luruper Chaussee 149, 22761 Hamburg, Germany \\
\vspace{.1cm}\\
2- Martin-Luther University Halle-Wittenberg, Department of Physics\\
Friedemann-Bach-Platz 6, 06108 Halle, Germany\\
\vspace{.1cm}\\
3- DESY-Hamburg site, Deutsches Elektronen-Synchrotron in der Helmholtz-Gemeinschaft\\
Notkestrasse 85, 22607 Hamburg, Germany\\
}

\maketitle

\begin{abstract}
In the ILD concept, a TPC is foreseen as the main tracking device. The MediTPC prototype has been developed at DESY to study the properties of Gas Electron Multipliers (GEMs) and to perform resolution studies. 
Studies of the influence of oxygen contaminations on the electron signal are presented here. Further, resolution results with new pad planes with \unit[$1.27 \times 7.0$]{$\mathrm{mm}^2$} pad pitch and in high magnetic fields are shown.
\end{abstract}

\section{Introduction}

The MediTPC prototype has been developed to study the feasibility of Gas Electron Multipliers as a gas amplification system for a high precision Time Projection Chamber (TPC) at the International Linear Collider (ILC).

The MediTPC has an overall length of \unit[80]{cm} (sensitive: \unit[66]{cm}) and a diameter of \unit[27]{cm}. With this prototype, several measurements have been performed in high magnetic fields of up to \unit[4]{T} at a test stand at DESY (see \cite{Diener:2006hk},\cite{Janssen:2008zz} and \cite{Linzmaier:2009zz}). For the gas amplification, a triple GEM setup has been used with a voltage of 320 to \unit[330]{V} per GEM. In contrast to previous measurements, two new pad planes ---one staggered and one non-staggered--- were used with a pad pitch of \unit[$1.27 \times 7.0$]{$\mathrm{mm}^2$} ($48 \times 11$ pads) instead of the previously used \unit[$2.2 \times 6.2$]{$\mathrm{mm}^2$} ($24 \times 6$ pads) pad planes.

The data reconstruction has been performed using the MultiFit software~\cite{Diener:2006hk},\cite{Janssen:2008zz}. In this program, two fit methods are implemented: the \emph{Chi Squared Fit Method} ---with the option to use external diffusion and defocussing information for the Pad Response Correction (PRC)--- and the \emph{Global Fit Method} which fits a likelihood function to the deposited charges. The latter has the option to use external diffusion information to stabilize the fit by calculating the charge cloud width instead of fitting it.

\section{Electron Attachment due to Oxygen Contamination}

Several data sets with different concentrations of oxygen contamination in the gas have been measured to study the influence of the contaminations on the electron charge signal. Drifting electrons can attach to oxygen impurities in the gas and so the signal will be weakened. This corresponds to a loss of primary statistics and hence to a degradation of the signal.
 
\begin{wrapfigure}{l}{0.52\columnwidth}
\centerline{\includegraphics[width=0.50\columnwidth]{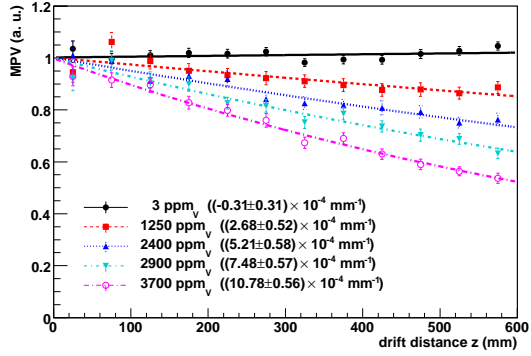}}
\caption{Measured mean hit charge for several oxygen concentrations in the gas.}\label{fig:MPV}
\vspace{-0.5cm}
\end{wrapfigure}
The number of free electrons $N$ in the gas is given by the following equation: 
\begin{equation}\begin{split} N(t) = exp( -At) \textnormal{, with:} \\
A = P(M) \times P(O_2) \times C_{O_2,M} \textnormal{,}
\label{eq:attco} \end{split} \end{equation}
where $A$ is called the \emph{attachment rate} and $t$ denotes the drift time. $P$ denotes the partial pressure of the chamber gas. $M$ denotes the gas mixture and O$_2$ its oxygen content. $C_{O_2,M}$ is called the \emph{attachment coefficient}.
As a measure for $N$, the Most Probable Value (MPV) of the Landau distributions of the hit charges (deposition per pad row) has been calculated at different drift lengths. The results shown in Figure \ref{fig:MPV} reflect the expectation. The signal decrease gets steeper at higher oxygen concentrations. But the influence of oxygen contaminations on the signal strength is only visible at rather high concentrations of more than several hundred $\textnormal{ppm}_\textnormal{V}$.

\begin{wraptable}{r}{0.36\columnwidth}
\centerline{\begin{tabular}{|c|c|}
\hline
O$_2$ [$\textnormal{ppm}_\textnormal{V}$] & A [$\mu\textnormal{s}^{-1}$ $\textnormal{bar}^{-2}$] \\ \hline
3 & 0 \\
1250 & 8.56 $\pm$ 1.92 \\
2400 & 8.66 $\pm$  1.08 \\
2900 & 10.31 $\pm$  0.90 \\
3700 & 11.64 $\pm$  0.68 \\\hline
\end{tabular}}
\caption{Attachment coefficient.}
\label{tab:attco}
\end{wraptable}
An exponential function has been fitted to the measurements to determine the attachment rate $A$ (see Equation \ref{eq:attco}). The results for different oxygen concentrations are shown in Table \ref{tab:attco}. The values are comparable to the results presented in~\cite{Huk:1987th}. Further, Garfield has been used to simulate these values, but a difference by a factor of about 100 has been observed between the measured and the simulated coefficients. The complete study of the electron attachment can be found in~\cite{Linzmaier:2009zz}.
 
\section{Point Resolution}

Besides the measurements mentioned above, several data sets have been measured to study the achievable point resolution at different magnetic fields up to \unit[4]{T}.
\begin{figure}[h!]
\centerline{
\includegraphics[width=0.50\columnwidth,clip]{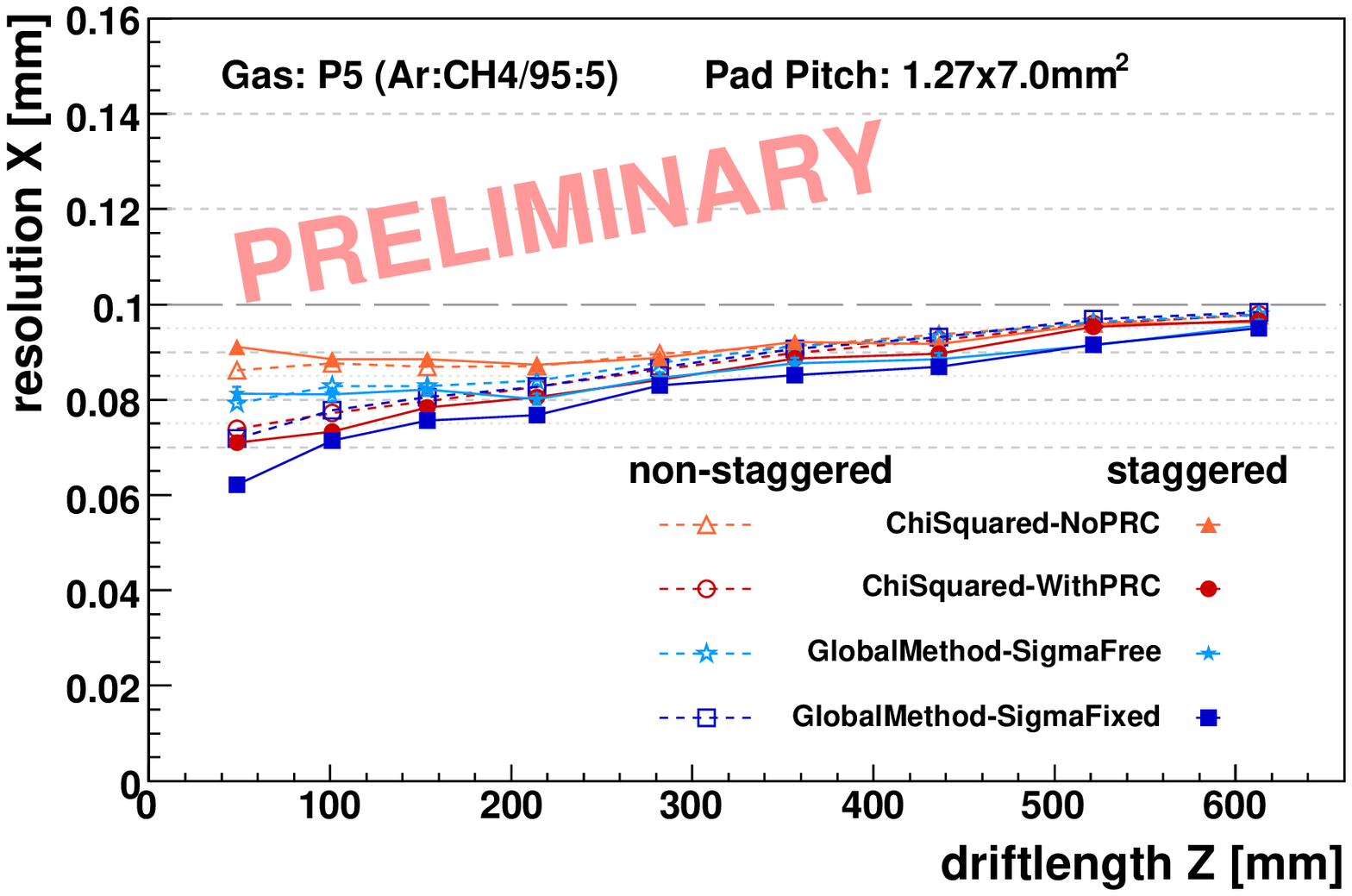}
\includegraphics[width=0.50\columnwidth,clip]{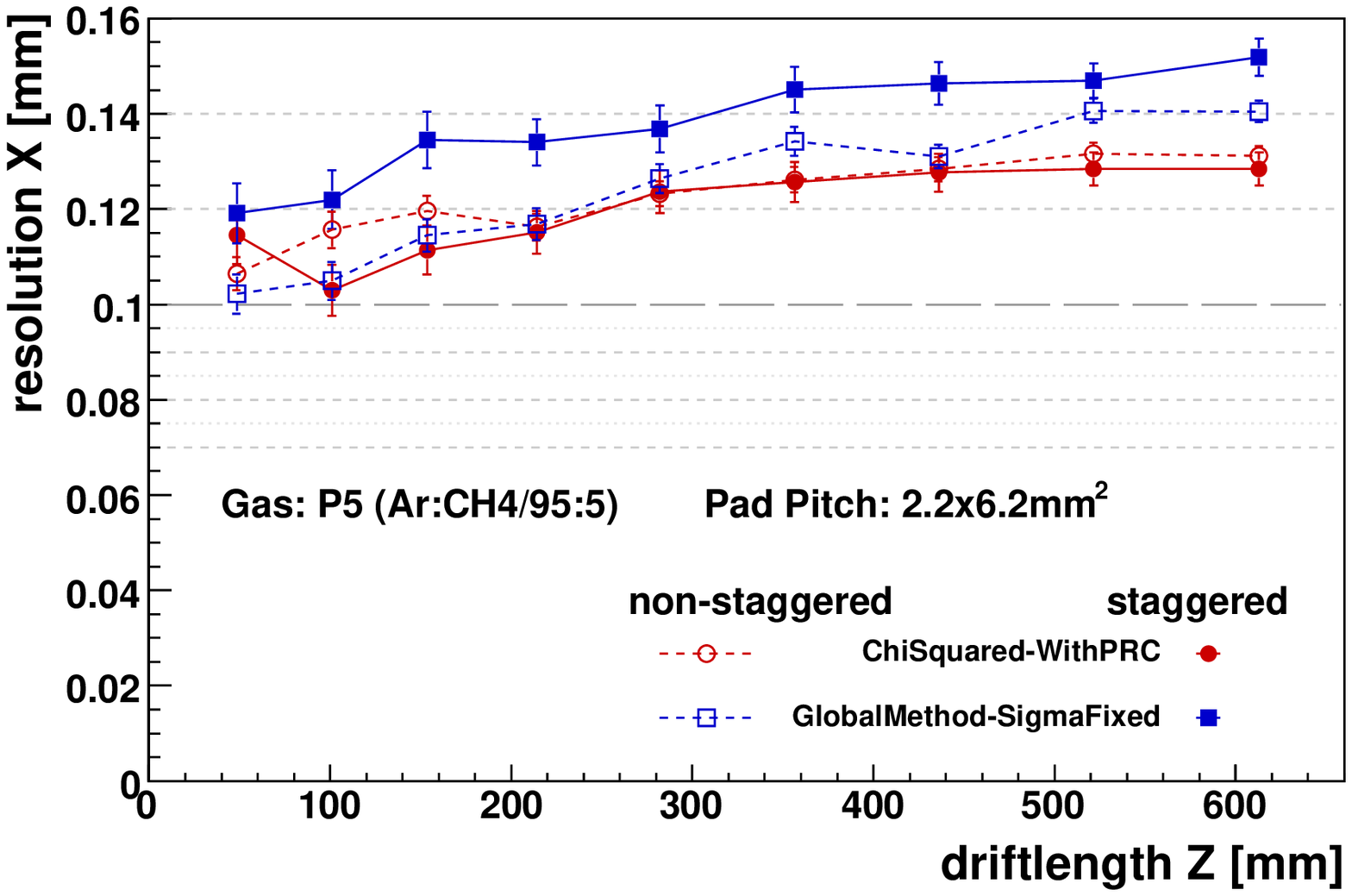}
}
\caption{Point resolution at \unit[4]{T}. Left side: small pads, right side: large pads~\cite{Janssen:2008zz}.}\label{Fig:Res4T}
\vspace{-0.1cm}
\end{figure}

The data sets have been reconstructed with both track fit methods and with and without using external diffusion information. The point resolution has been calculated using the so-called \emph{Geometric Mean Method}\footnote{Two residuals are being calculated for a track fit including the point and for a track fit without the point. The point resolution $\sigma$ is calculated from the geometric mean of the widths of both residual distributions.}.

\begin{wrapfigure}{r}{0.55\columnwidth}
\centerline{\includegraphics[width=0.54\columnwidth]{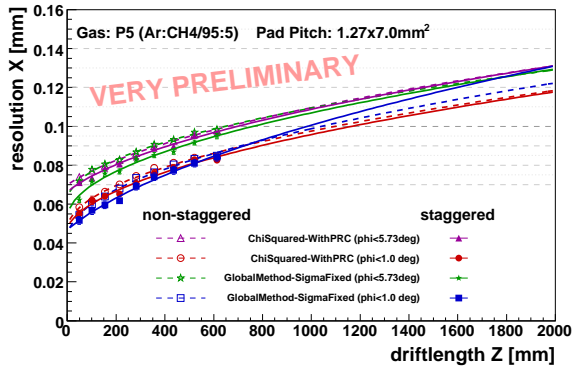}}
\caption{Extrapolation of resolution to a drift length of \unit[2]{m} for a 5.73$^{\circ}$ and a 1$^{\circ}$ track angle cut.}\label{Fig:ResExt}
\end{wrapfigure}
The results measured with a staggered and a non-staggered pad plane with a pad pitch of \unit[$1.27 \times 7.0$]{$\mathrm{mm}^2$} and 11 rows in a magnetic field of \unit[4]{T} are shown in the left plot of Figure \ref{Fig:Res4T}. For comparison, the results measured with a larger pad pitch of \unit[$2.2 \times 6.2$]{$\mathrm{mm}^2$} and 6 rows are shown in the right plot~\cite{Janssen:2008zz}. It is visible that a small pad pitch is necessary to reach the resolution goal of \unit[100]{$\mu$m}. Futher, it can be seen that with more rows the fit methods work more stable and the results of the different methods become comparable. Figure \ref{Fig:ResExt} shows the extrapolation of the results to a drift length of \unit[2]{m} for two angular cuts of 5.73$^{\circ}$ (=\unit[0.1]{rad}) and 1$^{\circ}$.

\section{Summary and Outlook}

The dependency of the electron signal on the attachment due to oxygen contaminations has been measured. A noticeable effect occurs only at oxygen concentrations of more than several \unit[100]{$\textnormal{ppm}_\textnormal{V}$}, which is well above the usual amount of about \unit[10]{$\textnormal{ppm}_\textnormal{V}$} during normal data taking. The huge discrepancy between the measured and the simulated values for the attachment coefficient will be studied.

The point resolution results at \unit[4]{T} indicate, that the resolution goal of less than \unit[100]{$\mu$m} over the whole drift length in a final TPC is in reach. A small pad size is essential to achieve this goal. The next steps include an optimization of the reconstruction ---especially regarding angular effects--- and an examination of the possibilities to limit the drift length dependent diffusion. If this diffusion could be limited, the resolution goal would be feasible over the two or more meters of drift length in the TPC of the final ILD detector.

We like to thank T.~Behnke, L.~Hallermann, M.~E.~Janssen, N.~Kanning, A.~Kaukher, K.~Komar, P.~Schade and O.~Sch{\"a}fer for their help.

\begin{footnotesize}

\end{footnotesize}

\end{document}